\documentclass[12pt,a4paper]{article}
\usepackage[latin9]{inputenc}
\usepackage{amssymb}

\makeatletter

\pdfpageheight\paperheight
\pdfpagewidth\paperwidth

\usepackage{amsfonts}

\makeatother

\begin{document}

\title{On the physics of the $so_{q}(4)$ hydrogen atom}

\author{P.\ G.\ Castro%
\thanks{{\em e-mail: paulo.castro@en.mar.mil.br}%
} ~and R.\ Kullock%
\thanks{{\em e-mail: ricardokl@cbpf.br}%
} \\
 \\
 \textit{$~^{\ast}$Escola Naval, } \\
\textit{Av.\ Almirante Sylvio de Noronha, Rio de Janeiro (RJ), Brazil} \\
 \\
 \textit{$~^{\dagger}$CBPF,} \\
\textit{R.\ Dr.\ Xavier Sigaud 150, Rio de Janeiro (RJ), Brazil} \vspace{0.5cm}
}
\maketitle
\begin{abstract}
In this work we investigate the $q$-deformation of the $so(4)$ dynamical
symmetry of the hydrogen atom using the theory of the quantum group
$su_{q}(2)$. We derive the energy spectrum in a physically consistent manner and find a degeneracy breaking as well as a
smaller Hilbert space. We point out that using the deformed Casimir as was done before leads to inconsistencies in the physical interpretation of the theory.

\vspace{0.2cm}

{\footnotesize { }\textbf{\footnotesize Keywords}{\footnotesize :
hydrogen atom, quantum group, dynamical symmetry}{\footnotesize \par}

 }{\footnotesize \par}
\end{abstract}
\newpage{}

\section{Introduction}

In this work we investigate the deformation of the
hydrogen atom using the $su_{q}(2)$ quantum group. Replacing the
original algebra with its deformed version, it is still possible to
use the canonical approach to find the energy levels, which now depend
on one extra quantum number. A new constraint arising as a consequence
of the orthogonality between the Runge-Lenz vector and the angular
momentum leads to a smaller Hilbert space. We also investigate more
closely the physical interpretation of the usual construction, using
the deformed Casimirs.

Quantum groups and deformations have gained a prominent role in physics,
with applications on a great variety of subjects, e.g.\ quantum field
theory on noncommutative spaces \cite{gr1,gr2}. They might have been
initially investigated in the context of the quantum Yang-Baxter equation
as a way of generating its solutions, but they are also the mathematical
tool used to investigate the idea of a minimal length \cite{fr1,fr2}.
For this reason, much attention has been devoted to looking into deformed
symmetries in physical systems, such as the deformed rotational symmetry
$su_{q}(2)$ of the harmonic oscillator \cite{ng}. The relation between $q$-deformation and nonlinearity has been discussed in \cite{manko}, where it is shown that the classical counterpart of a $q$-deformed oscillator has an amplitude-dependent frequency. In some references
the $q$-deformation is made at the level of the Schrödinger equation
\cite{micu,zhang}, in which case only the deformation of the rotational symmetry is treated. The hydrogen atom on a noncommutative curved spacetime has been recently investigated in \cite{vlad}.

Another possibility lies in the investigation of deformed dynamical
symmetries. The best known example is the relation between $so(4)$
and the spectrum of the hydrogen atom, achieved by using the Runge-Lenz
vector, which enlarges the symmetry group. The spectrum is then given
by the restriction of the representation $(i,j)$ to $(i,i)$. 

The
$q$-deformation of the dynamical symmetry of the hydrogen atom has
been attempted in many different ways,  such as the Kustaanheimo-Stiefel transformation \cite{KS,boiteux}, and the use of a Moyal-like $\star$-product \cite{rosa}. Other works have investigated this deformation through the separation
$so(4)=su(2)\oplus su(2)$ \cite{kn,sl,gora,yx}, but have overlooked an inconsistency in the physical interpretation of the results which shall be pointed out timely. 

The structure of this paper is as follows: we first briefly review
the derivation of the spectrum of the hydrogen atom using the $so(4)$
dynamical symmetry.  We  point out the inconsistencies in the physical interpretation of the deformation performed in previous works. We then present a physically consistent $q$-deformation of the dynamical symmetry of the hydrogen atom and derive 
the energy spectrum, showing the new dependence on the quantum numbers
and the reduction of the Hilbert space,  and discuss the
partial breaking of the degeneracy of the energy levels. We show that the same deformation parameter $q$ must be used on the two copies of $su(2)$.

\section{Undeformed case}
We start by briefly reviewing the derivation of the spectrum of the hydrogen atom using its dynamical symmetry $so(4)$ \cite{pauli, fock, bi}. 

We want to solve the Schr\"odinger equation for the Coulomb Hamiltonian
\begin{equation}
H=\frac{p^2}{2\mu}-\frac{k}{r},
\end{equation}
where $\mu$ is the reduced mass, $k = e^2/ 4 \pi \epsilon_0$ and $e$ the charge of the electron.

It is known from the Kepler problem that the vectors $\mathbf{L}=\mathbf{r}\times\mathbf{p}$ and $\mathbf{M}=\mathbf{v}\times\mathbf{L}-k\left(\frac{\mathbf{r}}{r}\right)$ are constants of motion. They are, respectively, the angular momentum and the Runge-Lenz vector and are, by construction, orthogonal to each other.

Now using the correspondence principle and taking the Hermitian part of $\mathbf{M}$, one finds that the commutation relations among $L_i$, $M_i$ and $H$ are
\begin{eqnarray}
\left[L_i,H\right]&=&\left[M_i,H\right]=0\\
\left[L_i,L_j\right]&=&i\hbar\epsilon_{ijk}L_k\\
\left[L_i,M_j\right]&=&i\hbar\epsilon_{ijk}M_k\\
\left[M_i,M_j\right]&=&-\frac{2i\hbar}{\mu}\epsilon_{ijk}L_kH,
\end{eqnarray}
and, from the commutation relations between $x_i$ and $p_i$, that the relations
\begin{eqnarray}
\mathbf{L}\cdot\mathbf{M}=\mathbf{M}\cdot\mathbf{L}=0,\label{c1}\\
\mathbf{M}^2=k^2+\frac{2}{\mu}(\mathbf{L}^2+\hbar^2)H\label{c2}
\end{eqnarray}
hold.

If we now restrict ourselves to the Hilbert subspace corresponding to a negative eigenvalue $E$ of $H$, we can rescale $\mathbf{M}$ as 
\begin{equation}
\tilde{\mathbf{M}}=\sqrt{-\frac{\mu}{2E}}\mathbf{M},
\end{equation}
and therefore introduce the operators 
\begin{eqnarray}
\mathbf{I}&=&\frac{\mathbf{L}+\tilde{\mathbf{M}}}{2}\\
\mathbf{J}&=&\frac{\mathbf{L}-\tilde{\mathbf{M}}}{2}
\end{eqnarray}
which form two disjoint sets of operators satisfying the $su(2)$ algebra: 
\begin{eqnarray}
 \left[I_z,I_{\pm}\right] = I_{\pm}, && \left[I_+,I_-\right] = 2I_z \\
 \left[J_z,J_{\pm}\right] = J_{\pm}, && \left[J_+,J_-\right] = 2J_z,
\end{eqnarray}
so that we have in principle 4 quantum numbers, $i,m$ and $j,p$, diagonalizing $\mathbf{I}^2, I_z$
and $\mathbf{J}^2, J_z$, respectively.

Relating these 2 algebras to the hydrogen atom means rewriting relations (\ref{c1}) and (\ref{c2}) in terms of $\mathbf{I}$ and $\mathbf{J}$ as
\begin{equation}\label{c3}
 (\mathbf{I}^2 - \mathbf{J}^2) |i,m,j,p\rangle = 0,
\end{equation}
and
\begin{equation}\label{c4}
 (\mathbf{I}^2 + \mathbf{J}^2) |i,m,j,p\rangle = - \left( \frac{ \mu k^2}{4E} + \frac{\hbar^2}{2}
\right)|i,m,j,p\rangle.
\end{equation}

Since
\begin{eqnarray}
 \mathbf{I}^2 |i,m,j,p\rangle &=& i(i+1)\hbar^2 |i,m,j,p\rangle \\
 \mathbf{J}^2 |i,m,j,p\rangle &=& j(j+1)\hbar^2 |i,m,j,p\rangle,
\end{eqnarray}
relation (\ref{c3}) yields that $i=j$, and, as a result of (\ref{c4}), we have that 
\begin{equation}
 2\hbar^2j(j+1) = - \left( \frac{ \mu k^2}{4E} + \frac{\hbar^2}{2}\right),
\end{equation}
leading to
\begin{equation}
 E = \frac{-\mu k^2}{2\hbar^2(2j +1)^2},
\end{equation}
so that we can identify $2j+1 = n$ with the principal quantum number of the hydrogen atom.

Since $i=j$, the Hilbert space is of the appropriate size.

It is important to note that $j$ can take half-integer values, and that $I_z$ and $J_z$ can each have $2j+1$ independent eigenvalues, so that the degeneracy of a given state is $(2j+1)^2=n^2$ as expected.

\section{The $q$-deformed $so(4)$}

We choose to deform the $so(4)=su(2)\oplus su(2)$ dynamical symmetry of the hydrogen atom by a parameter
$q$ by means of the well-established theory of the quantum group $su_{q}(2)$ \cite{mac,bied,sf,kd},
where the commutation relations are written as 
\begin{eqnarray}
\left[I_{z},I_{\pm}\right]=I_{\pm}, &  & \left[I_{+},I_{-}\right]=[2I_{z}]\\
\left[J_{z},J_{\pm}\right]=J_{\pm}, &  & \left[J_{+},J_{-}\right]=[2J_{z}],
\end{eqnarray}
 where 
\begin{equation}
[x]=\frac{q^{x}-q^{-x}}{q-q^{-1}}=\frac{\sinh(sx)}{\sinh s},
\end{equation}
 with $s=\ln{q}$ and $q$ a real parameter. In this section, we take
$\hbar=1$. Notice that in priniciple we could have chosen two different parameters,
one for each copy of $su(2)$. This will be discussed further ahead.

As for the representation, this means that $I_z$ and $J_z$ will act on the Hilbert space exactly
as before, while for the other operators we have
\begin{equation}\label{r1}
 I_{\pm} |i,m,j,p\rangle = \sqrt{[i \pm m +1][i \mp m]} |i,m\pm 1,j,p\rangle,
\end{equation}
and analogously for $J_{\pm}$, so that relations such as
\begin{eqnarray}\label{r2}
 I_\pm I_\mp |i,m,j,p\rangle = ([i][i+1] - [m][m\mp 1] )|i,m,j,p\rangle ,
\end{eqnarray}
are still diagonal in the old Hilbert space (i.e., using the same basis).

As $I_{\pm}$ and $J_{\pm}$ have similar
commutation relations with $I_{z}$ and $J_{z}$, they raise and lower their
eigenvalues exactly like before. This means that the Hilbert space will
have the same quantum numbers, one related to the Casimir and one to the $z$-component of the angular momentum.

\subsection{Using the deformed Casimirs}

 The center of the algebra is  changed upon deformation of the commutation relations. This means that if we are to merely follow the formal aspects of the derivation of the spectrum using $so(4)$, as was done in aforecited earlier papers, we should take the Casimirs of the deformed algebra, given by
\begin{eqnarray}
C_{2}^{I}|i,m,j,p\rangle=\left(\frac{1}{2}(I_{+}I_{-}+I_{-}I_{+})+\frac{[2]}{2}[I_{z}]^{2}\right)|i,m,j,p\rangle=[i][i+1]|i,m,j,p\rangle, \nonumber \\
C_{2}^{J}|i,m,j,p\rangle=\left(\frac{1}{2}(J_{+}J_{-}+J_{-}J_{+})+\frac{[2]}{2}[J_{z}]^{2}\right)|i,m,j,p\rangle=[j][j+1]|i,m,j,p\rangle.\nonumber \\
 { } ~
\end{eqnarray}

Now the orthogonality condition should read 
\begin{equation}
C_{2}^{I}|i,m,j,p\rangle=C_{2}^{J}|i,m,j,p\rangle,
\end{equation}
leading only to $i=j$, so that the Hilbert space has the same size
as in the undeformed case. The energy in this setting becomes 

\begin{equation}
E_{j}=\frac{-\mu k^{2}}{8[j][j+1]}
\end{equation}
and we have the same degeneracy as before.

Notice, however, that while the generators of the two $su_{q}(2)$ have
the same interpretation as in the undeformed case, the equations
defining the spectrum will have a different meaning. The orthogonality condition
will now read
\begin{equation}
L_{x}\tilde{M}_{x}+L_{y}\tilde{M}_{y}+\frac{[2]}{2}[L_{z}][\tilde{M}_{z}]=0
\end{equation}
or, more explicitly
\begin{equation}
L_{x}\tilde{M}_{x}+L_{y}\tilde{M}_{y}+\frac{\sinh(2s)\sinh(sL_{z})\sinh(s\tilde{M}_{z})}{2\sinh^{3}s}=0.
\end{equation}

This means that
the angular momentum and the Runge-Lenz vector are no longer orthogonal. To see that, one could take the first
elements in the $\sinh$ expansion and regroup them as a scalar product,  and all the remaining nonvanishing terms will make the breaking of the orthogonality explicit. This leads to an inconsistent situation,
because the Runge-Lenz vector orthogonality to the angular momentum
in the undeformed case comes directly from its definition as a cross
product.

Similarly, the expression for the energy is modified by using the deformed
Casimir, so that the actual Hamiltonian described by this $q$-deformed system is
different from the original one, although the current approach makes
it difficult to notice this fact.

The expression for energy is implicitly given by equation (\ref{c2}). In the undeformed case, solving this expression leads to the Coulomb potential. In the $q$-deformed case, the implicit expression reads
\begin{eqnarray}
\nonumber L_x^2+L_y^2 &-&\frac{\mu}{2E}\left(M_x^2+M_y^2\right)+\frac{\sinh(2s)}{2\sinh ^3 s}\left\{\sinh^2\left(s\left(L_z+\sqrt{-\frac{\mu}{2E}}M_z\right)\right)\right.+  \\
 &+&\left.\sinh^2\left(s\left(L_z-\sqrt{-\frac{\mu}{2E}}M_z\right)\right)\right\}=-\frac{\mu k^2}{4E}+\frac{\hbar^2}{2}.
\end{eqnarray}

If we were to solve this equation for $E$, we would no longer have the Coulomb potential. Notice that all the problematic terms involve only the $z$-direction, indicating that the system is no longer described by a central potential.

What we can gather from all these remarks is that the route for the $q$-deformation of the hydrogen atom pursued in previous works has flaws and inconsistencies regarding its physical interpretation, the most prominent of which are the loss of orthogonality between the angular momentum and the Runge-Lenz vector and the fact that the $q$-deformed Hamiltonian cannot be interpreted as related to the classical Coulomb Hamiltonian either.  The latter situation is somewhat analogous to \cite{manko}, where the $q$-deformed harmonic oscillator is actually nonlinear.

In the next section we shall present a way to circumvent these problems and introduce a $q$-deformation of the hydrogen atom that is physically consistent and gives rise to interesting new features.

\subsection{The relation $\mathbf{L}\cdot\mathbf{M}=0$ and the energy shell}

We now consider the relations (\ref{c3}) and (\ref{c4}). They are the Casimirs of the undeformed $so(4)$ algebra, and, moreover, (\ref{c3}) encodes the condition of orthogonality between  $\mathbf{L}$ and $\mathbf{M}$. 

Using expressions (\ref{r1}) and (\ref{r2}), it is easy to deform the relation (\ref{c3})  and see that $(\mathbf{I}^2 - \mathbf{J}^2) |i,m,j,p\rangle = 0$ implies that $i=j$ and $m^2 = p^2$. Note that the second restriction is completely new
to the deformed case. This means that the resulting Hilbert space will be smaller, and
that there will be a jump in its size when we the deformation parameter
goes back to $q=1$. This resembles a ``phase transition''.

Investigating relation (\ref{c4}) we find that it is still diagonal
-- even though it is not the Casimir of the deformed algebra -- and,
by imposing the orthogonality condition, we find that
\begin{equation}
(\mathbf{I}^{2}+\mathbf{J}^{2})|i,m,j,p\rangle=\left(2[j][j+1]-[m]([m+1]+[m-1])+2m^{2}\right)|i,m,j,p\rangle.
\end{equation}

This immediately gives the deformed energy for the hydrogen atom as
\begin{equation}
E_{jm}=\frac{-\mu k^{2}}{8[j][j+1]-4[m]([m+1]+[m-1])+8m^{2}+2}.
\end{equation}

It easy to realize that this has the correct limit when $q\rightarrow1$.

Since the energy $E_{jm}$ now depends additionaly on the quantum
number $m$, part of the degeneracy of the energy spectrum is broken.
It is easy to see that changing $m\rightarrow-m$ leaves the energy
invariant, and thus for each $j$ there are $j+1$ possible energy
levels.

Let us now work out the degeneracy of each of these levels. Because
the result does not depend on $p$, one could expect the usual $2j+1$
degeneracy that comes from it, but this is not the case due to the
constraint $(\mathbf{I}^{2}-\mathbf{J}^{2})|i,m,j,p\rangle=0$, which
makes $m^{2}=p^{2}$ and thus $p=\pm m$. This means that $E_{jm}$
is four-fold degenerate for $m\neq0$ and nondegenerate for $m=0$,
and therefore each value of $j$ corresponds to $4j+1$ states, as
opposed to the usual $(2j+1)^{2}$.

It is interesting to note that the degeneracy of the hydrogen spectrum in noncommutative settings is a much investigated issue. Depending on the construction that is used, it can be totally \cite{cha} or partially \cite{vlad} broken, or even totally retained \cite{pres}.

\subsection{Different deformation parameters}

Since we have two copies of the $su(2)$, we could be tempted to deform them
with different parameters $q$ and $q'$. In this case, the orthogonality
condition (using the deformed Casimir) will give us
\begin{equation}
\frac{\sinh(si)\sinh(s(i+1))}{\sinh^{2}(s)}=\frac{\sinh(tj)\sinh(t(j+1))}{\sinh^{2}(t)},
\end{equation}
with $s=\ln{q}$ and $t=\ln{q'}$. The leading order here will give
$i(i+1)=j(j+1)$, just like the undeformed case, so that $i=j$. Using
the next-to-leading order we arrive at the conclusion that
$s=t$, so $q=q'$. In other words, if we try to use two different
deformation parameters then the deformed orthogonality condition states that
they should be the same.

In the novel approach using undeformed Casimirs the situation is analogous. The first order implies $i=j$ and the second order either leads to $q=q'$ (as before) or, for very particular values of the ratio $s/t$, to a single-eigenstate Hilbert space. The second case is not worth investigating. 

This means that in both approaches we are forced to use the same deformation parameter on both copies of $su_q(2)$.

\section{Conclusion}

In this work we constructed a physically consistent $q$-deformed version of the energy
spectrum of the hydrogen atom by deforming the commutation relations
of the $so(4)$ dynamical symmetry using the $su_{q}(2)$ quantum
group. In this setting, the actions of $I_{\pm}$ and $J_{\pm}$ are analogous
to the undeformed ones, with the coefficients replaced by $q$-numbers.

The resulting Hilbert space is smaller, and the degeneracy of
the energy spectrum is partially broken, due to the fact that, although
$\mathbf{I}^{2}$ and $\mathbf{J}^{2}$ are still diagonal, their
eigenvalues have an extra dependence on $m$ and $p$. The degeneracy
is not completely removed because of the constraint $m=\pm p$ and
the invariance of the energy under $m\rightarrow-m$. The residual
degeneracy does not depend on $j$. The number of states corresponding
to each value of the quantum number $j$ is smaller than in the usual
case.

We can compare the spectrum and Hilbert space
found by our approach, i.e.\ using the orthogonality and energy relations, to
the spectrum obtained using the deformed Casimirs,
where the Hilbert space has the same size as in the
undeformed case and the energy degeneracy is untouched, but the physical
interpretation becomes more complicated, with a Runge-Lenz vector
that is not orthogonal to the angular momentum. It should be noted
that the Runge-Lenz vector is still a conserved quantity. The meaning
of the Hamiltonian is also modified but this is trickier to see, because  the $so(4)$ symmetry is invoked only after
a restriction to an energy shell. It could be argued from these remarks that the $q$-deformed construction presented in earlier works is not really related to the classical Coulomb Hamiltonian, and thus physically invalid even if mathematically correct.

 We point out that it is not possible to use different deformation parameters on each of the two $su(2)$ copies that make up $so(4)$.

{}~ \\
{}~ 
\textbf{\large Acknowledgments}{} ~\\
{}~

R.K.\ acknowledges support by CNPq -- Brasil. This work was partially developed at Universit\"at Wien (R.K.). Both authors are thankful to F.\ Toppan and  V.\ Kupriyanov for enlightening comments.

\end{document}